\documentclass[aps,twocolumn,notitlepage,superscriptaddress,floatfix]{revtex4}

\usepackage{amssymb,amsfonts,amsmath}
\usepackage[latin1,utf8]{inputenc}
\usepackage{hyperref}
\usepackage{graphicx}

\begin{document}

\title{The regular black hole in four dimensional Born-Infeld gravity}

\author{Christian G. B\"{o}hmer}
\email{c.boehmer@ucl.ac.uk}
\affiliation{Department of Mathematics, University College London, Gower Street, London WC1E 6BT, United Kingdom.}

\author{Franco Fiorini}
\email{francof@cab.cnea.gov.ar}
\affiliation{Departamento de Ingeniería en Telecomunicaciones and Instituto Balseiro, Centro Atómico Bariloche, Av. Ezequiel Bustillo 9500, CP8400, S.C. de Bariloche, Río Negro, Argentina.}

\date{\today}

\begin{abstract}
In the context of Born-Infeld gravity theories we report the existence of a regular black hole interior representing a spherically symmetric vacuum solution of the theory. It reduces to the Schwarzschild interior metric in the weak field region. In particular, there is a new length scale which is related to the Born-Infeld parameter $\lambda$. This endows the spacetime with an inner (i.e.~well inside the event horizon) asymptotic region which is unattainable for observers. The central curvature singularity is replaced by an infinitely long cosmic string with constant curvature invariants related to $\lambda$. The presence of this limiting curvature spacetime renders the black hole timelike and null geodesically complete, free from the classical Schwarzschild singularity. The transition between the usual black hole interior and this maximum curvature space is achieved without introducing any kind of matter content nor topological changes.
\end{abstract}

\maketitle

\section{Introduction}

The history and development of black holes physics represent one of the most intriguing and profound stories of modern science. Born as mere curiosities out of Einstein's field equations during the very early days of General Relativity (GR), black holes have gradually evolved towards more physical grounds, being today a matter of study not only in the context of astrophysics and cosmology, but also in analogue systems as fluid dynamics and optics~\cite{Analogue1,Analogue2,Analogue3}. Chandrasekhar's famous work~\cite{Chandra} on the maximum mass of white dwarfs pointed the way to our present-day picture that, for bodies of too large a mass, concentrated in too small a volume, unstoppable collapse will ensue, leading to a singularity in the structure of space-time. The details behind this continued gravitational collapse have been worked out for the first time by Oppenheimer and Snyder~\cite{Oppen} who discovered that a radial non-rotating dust cloud will contract indefinitely towards a point with infinite energy density and pressure. This result constitutes the first example of an intrinsic strong curvature singularity within black hole physics.

Subsequent pioneering developments by Penrose and Hawking~\cite{Sin1,Sin2,Sin3} in particular, led to our present understanding that the existence of some form of singularity is, in fact, a characteristic feature of large classes of solutions of the field equations, not only the property of the final state of collapsing stars. In the context of spherically symmetric general relativistic solutions, the quest for geodesically complete spacetimes has been initiated since Bardeen~\cite{Bardeen} discovered a regular black hole obeying the weak energy condition, which was influential in leading the direction of subsequent research. As was pointed out in~\cite{Borde}, many of these Bardeen-like black holes that came after the original proposal owe their regularity to topological changes, which offers the possibility to obtain spaces with a maximum curvature inside the black hole~\cite{Markov1,Markov2,Frolov,Brand1}. This feature, of course, involves the introduction of matter fields in a rather ad hoc manner. Other regular black hole models were also proposed by coupling Einstein's theory with nonlinear electrodynamics~\cite{Eloy}. Nevertheless, most of these regular models constitute solutions of Einstein's equations with sources whose physical interpretation, even though satisfying reasonable (weak) energy conditions, are quite obscure and not related with any particular experimentally suggested model for the matter or electromagnetic fields, see also~\cite{Bronnikov:2001} for related views. Regular black hole interior solutions were also found in Loop Quantum Gravity, see~\cite{Bohmer:2007}, one of the promising candidates for a theory of quantum gravity.

Nowadays, black hole physics experiences a remarkable revival not only because of the extraordinary observational evidence concerning the existence of these bizarre objects in the Universe (including the one in our galactic centre~\cite{Ghez}), and the emission of gravitational waves coming from a binary black hole mergers~\cite{Ligo}, but also in relation to discussions on more fundamental grounds, as black hole entropy and the information loss paradox~\cite{HP}. However, despite all these technical advances in the area, the physics close to the singularity, where strong curvature effects rule the fate of infalling observers, is largely unknown.

The purpose of this work is to inquire into this matter more closely by exhibiting what seems to be the first regular, vacuum black hole interior which appears as a solution of the so called Born-Infeld (BI) gravity, a theory which already has shown non singular states in several cosmological contexts~\cite{FF2007}. This regular spacetime owes its geodesic completeness to a purely geometrical effects by virtue of the fact that gravity becomes repulsive in the strong field regime, and not by the inclusion of matter fields or topological changes. Throughout the paper, we will adopt the signature $-2$, and, as usual, Latin indexes $a:(0),(1),\ldots$ refer to tangent-space objects while Greek $\mu:0,1,\ldots$ denote spacetime components.

\section{On Born-Infeld gravity and the Schwarzschild interior}

It was known since the 1960s that the interior region of the Schwarzschild black hole can be viewed as a special homogeneous but anisotropic cosmological manifold $\mathcal{M}$ with topology $\mathbb{R} \times \mathbb{R} \times \mathbb{S}^2$. This Kantowski-Sachs (KS) equivalence~\cite{KS66} shows that the metric of the interior region can be written as
\begin{align}
  ds^2 = dt^2-b^2(t)\,dz^2-a^2(t)\,d\Omega^2,
  \label{metrica}
\end{align}
where $a(t)$ and $b(t)$ are the scale factors depending only on the proper time and $d\Omega^2= d\theta^2+ \sin^2\negmedspace\theta d\phi^2$ is the line element of the 2-sphere $\mathbb{S}^2$. The scale factors are obtained by solving the two independent vacuum Einstein field equations for metric~(\ref{metrica}), namely
\begin{align}
  a^{-2}+H_{a}^{2}+2H_{a}H_{b} &= 0
  \nonumber \\
  a^{-2}+3H_{a}^{2}+2\dot{H}_{a} &= 0,
  \label{sisrg}
\end{align}
where $H_{a}=\dot{a}/a$ and $H_{b}=\dot{b}/b$ are the Hubble factors and the dot denotes differentiation with respect to time. In terms of the new time function $\eta(t)$ defined implicitly by $t-t_{0} =a_{1} (\eta+\sin\eta\cos\eta)$, the solution of the system~(\ref{sisrg}) is given by
\begin{align}
  b(t)=b_{1}\,\tan\left(\eta(t)\right),\quad
  a(t) = a_{1}\,\cos\negmedspace^2\left(\eta(t)\right).
  \label{func-a}
\end{align}
The integration constants $a_{1}$ and $b_{1}$ satisfy $-\pi/2\leq b_{1}<0$ and $a_{1} \neq 0$. According to this view, the space-time singularity corresponds to $\eta(t)=\pi/2$, where the two-spheres collapse to a point, and the laws of physics simply cease to exist. Details of the KS interior can be found in~\cite{Dor08}.

Born-Infeld gravity formulated in Weitzenb\"{o}ck space lies within the context of $f(T)$ gravity, which can be viewed as a natural extension of Einstein GR in its absolute parallelism (or teleparallel) form. In general, $f(T)$ gravitational theories are ruled by the action in four spacetime dimensions~\cite{RafaBen}
\begin{align}
  I=\frac{1}{16 \pi G }\int f(T)\,  e\, d^4x+I_{\rm matter},
  \label{actionfT3D0}
\end{align}%
where $e=\sqrt{|\det g_{\mu\nu}|}$. The Weitzenb\"{o}ck scalar
\begin{align}
  T = S^{a}{}_{\mu\nu} T_{a}{}^{\mu\nu},
  \label{Weitinvar}
\end{align}
is constructed quadratically from the torsion tensor $T^{a}_{\,\,\mu\nu}=\partial_{\mu} e^{a}_{\nu}-\partial_{\nu} e^{a}_{\mu}$ by means of the \emph{superpotential} $S^{a}_{\,\,\mu \nu}$
\begin{multline}
  S^{a}{}_{\mu\nu} = \frac{1}{4}
  (T^{a}{}_{\mu\nu} - T_{\mu \nu }{}^{a} + T_{\nu \mu }{}^{a}) +
  \frac{1}{2} (\delta_{\mu}^{a} T_{\sigma\nu}{}^{\sigma} -
  \delta_{\nu}^{a} T_{\sigma\mu}{}^{\sigma}).
  \label{tensorS}
\end{multline}
Of course GR is contained in~(\ref{actionfT3D0}) when $f(T)=T$, because $T =-R + 2 e^{-1} \partial_{\nu }(e\,T_{\sigma }{}^{\sigma\nu})$, so the the Weitzenb\"{o}ck scalar differs from the scalar curvature $R$ in the Hilbert-Einstein action, by a total derivative term, see for instance~\cite{TEGRbook} The dynamical equations in $f(T)$ theories, for matter coupled to the metric in the usual way, are obtained by varying action~(\ref{actionfT3D0}) with respect to the components of the vierbein field $e^{a}_{\mu}$
\begin{multline}
  \bigl(e^{-1}\partial_\mu(e\ S_a{}^{\mu\nu})+e_a^\lambda T^\rho{}_{\mu\lambda} S_\rho{}^{\mu\nu}\bigr)
  f^{\prime } \\ +
  S_a{}^{\mu\nu} \partial_\mu T f^{\prime \prime } - \frac{1}{4}
  e_a^\nu f = -4\pi G e_a^\lambda \mathcal{T}_\lambda{}^{\nu},
\label{ecuaciones}
\end{multline}
where the prime means differentiation with respect to $T$ and $\mathcal{T}_\lambda{}^{\nu}$ is the energy-momentum tensor coming from $I_{\rm matter}$.

The additional degrees of freedom arising from the lack of local Lorentz invariance in the $f(T)$ framework~\cite{RafaMajo}, pre-establish a global frame which encodes the geometry by means of a parallelization of the spacetime in consideration. Generally it is difficult, for a given geometrical setting, to construct proper tetrad fields as no systematic construction procedure exists.

In other words, the equations of motion~(\ref{ecuaciones}) determine the entire tetrad field, and not only the ten components involved in the definition of the metric tensor $g_{\mu\nu}=e^{a}_{\mu}e^{b}_{\nu}\eta_{ab}$. Nonetheless this \emph{tetrad grid} is non-unique, because a partial local Lorentz invariance persists in $f(T)$ gravity through the remnant group of symmetries~\cite{nos2015}. It is worth mentioning that a covariant formulation of teleparallel gravity models has recently been proposed~\cite{Krssak:2015oua}.

A set of adequate tetrads for the spacetime $\mathcal{M}$ with metric~(\ref{metrica}) was found in Ref.~\cite{Fio14} in the context of cosmological models with additional compact dimensions. The cosmological spacetime $\mathcal{M}$ can be foliated by Cauchy hypersurfaces with topology $\mathcal{M}_{3}= \mathbb{R} \times \mathbb{S}^{2}$ by means of the global time function (proper time $t$), and a parallelization can be found for $\mathcal{M}_{3}$. Then, the tetrad inherits the factor structure $\mathcal{M}= \mathbb{R} \times \mathcal{M}_{3}$, and it reads
\begin{align}
  \label{parasixs2}
  e^{(t)} &= dt,\qquad
  e^{(z)} = b(t)\cos\theta \,dz+a(t) \sin^2\theta \,d\phi, \\
  e^{(\theta)} &= \sin\phi [b(t)\sin\theta dz+a(t)(\cot\phi d\theta-\sin\theta \cos\theta d\phi)], \nonumber \\
  e^{(\phi)} &= \cos\phi [b(t)\sin\theta dz-a(t)(\tan\phi d\theta+\sin\theta\cos\theta d\phi)].\nonumber
\end{align}
Notice that these fields are highly non-trivial due to the topological structure of $\mathcal{M}$, and they are all far from being simply the square root of the diagonal metric~(\ref{metrica}). Using the frame~(\ref{parasixs2}) we can compute the Weitzenb\"{o}ck invariant~(\ref{Weitinvar}) which becomes
\begin{align}
  T=-2(-a^{-2}+H_{a}^2)-4 H_{a}H_{b},
  \label{inv}
\end{align}
which, in the case of the Schwarzschild solution in Kantowski-Sachs form, Eqs.~(\ref{func-a}), reduces to
\begin{align}
 T_{\rm KS}=4\,a^{-2}=4\,a_{1}^{-2}\, \cos^{-4}(\eta(t)) \ .
 \label{inveva}
\end{align}
This is a very interesting result. Unlike in the exterior Schwarzschild geometry, $T_{\rm KS}$ is everywhere non-null for a proper frame in the interior region of the black hole. This fact opens the door to a potential deformation of the interior spacetime~\cite{FF11}. In order to show that this is indeed the case, we proceed now to write the dynamical equations for an arbitrary vacuum $f(T)$ theory using tetrad~(\ref{parasixs2}).

The two independent field equations are given by
\begin{align}
  f+4f'(H_{a}^2+2H_{a}H_{b}) & =0,
  \label{Friedmann} \\
  f''H_{a}\dot{T}+f'(H_{a}H_{b}+2H_{a}^{2}+\dot{H}_{a})+\frac{f}{4} &=0,
  \label{sec-esp-1}
\end{align}
where the former is the Friedmann-like constraint. Let us note that in the case of GR ($f=T$, $f'=1$, $f''=0$, and using the general expression for $T$ given in~(\ref{inv})), one re-obtains the KS equations~(\ref{sisrg}).

Even though Born-Infeld like schemes for the gravitational field formulated in the (usual) Riemannian context can be traced back to the late 1990s \cite{Deser}, it was shown about a decade ago that Born-Infeld models in Weitzenb\"{o}ck space were particularly convincing at the time of trying to answer long standing questions of fundamental character in gravitational physics, as the one concerning the singularity at the origin of the Universe~\cite{FF2007}. Similarly, more general BI gravitational theories based on absolute parallelism were considered more recently, in which regular bouncing cosmological solutions where obtained by purely geometrical means, i.e.~without invoking any sort of unconventional matter content~\cite{yo2013,revOlmo}. We go back to the original $f(T)$-BI like proposal introduced in~\cite{FF2007}, and consider the ultraviolet deformation given by
\begin{align}
  f(T)=\lambda[\sqrt{1+2T/\lambda}-1],
  \label{BI}
\end{align}
which recovers Einstein's gravity in regions where $T/\lambda \ll 1$, $\lambda$ being the BI constant. This BI constant has units of inverse length squared. More explicitly, we have $f(T) = T - T^2/(2\lambda) + O(1/\lambda^2)$ which means that $\lambda$ introduces the length scale $|\lambda|^{-1/2}$ at which local Lorentz invariance would not longer hold~\cite{FF2011PLB}. 

An additional motivation for the form of the function (\ref{BI}) can be motivated as follows, see~\cite{yo2013,09104693}. Let us begin with the free particle Lagrangian $L_p=m\dot{x}^2/2$ in Newtonian mechanics. If we write its Born-Infeld analogue as $L_{\rm BI} = \sqrt{1+L_p/(mc^2)} - 1$ we introduce a new scale at which the theory changes. This Lagrangian is that of a relativistic particle. In this case, for energies much smaller that $mc^2$ one recovers Newtonian mechanics.

The combination of equations~(\ref{inv}) and~(\ref{Friedmann}), for the BI function (\ref{BI}) leads to
\begin{align}
  H_{a}^2+2H_{a}H_{b}=-\frac{1}{a^2(t)}\Big(1+\frac{4}{\lambda a^2(t)}\Big).
  \label{CombiBI}
\end{align}
This expression is very enlightening. If $\lambda=-|\lambda|$, a static ($H_{a}=0$) solution exists when $a_{c}=2 |\lambda|^{-1/2}$, foretelling the evolution of the scale factor $a(t)$ towards a \emph{critical} radius $a_{c}$, where the 2-spheres of the interior metric~(\ref{metrica}) with constant $t$ and $z$ tend to a static configuration. This has no GR anaologue when setting $\lambda\rightarrow\infty$ in~(\ref{CombiBI}). When the scale factor approaches the Schwarzschild singularity as $a(t)\rightarrow 0$, one finds $T\rightarrow\infty$. The functional form of $T$ in the BI case is easily obtained by combining~(\ref{CombiBI}) and~(\ref{inv}), namely
\begin{align}
  T_{\rm BI}=\frac{4}{a^2(t)}\Big(1-\frac{2}{|\lambda| a^2(t)}\Big), \label{invparaBI}
\end{align}
which reaches the maximum value $T_{\rm max}=|\lambda|/2$ when $a=a_{c}$.

\section{The regular Black Hole interior}

In order to get a differential equation for the scale factor $a(t)$, we can proceed as follows: Beginning with Eq.~(\ref{CombiBI}) we can obtain $H_{b}$ as a function of $a(t)$ and $H_{a}$. With this information and the Friedmann-like equation~(\ref{Friedmann}), we can substitute into~(\ref{sec-esp-1}) using the BI function~(\ref{BI}) and the Weitzenb\"{o}ck invariant~(\ref{invparaBI}). After some algebra we obtain
\begin{align}
  \Big[1-\frac{4}{|\lambda| a^{2}}\Big]
  \Big[\frac{1}{a^{2}}\Big(1-\frac{4}{|\lambda| a^{2}}\Big)+3 H_{a}^{2}+2\dot{H}_{a}\Big] -
  \frac{16 H^{2}_{a}}{|\lambda| a^{2}}=0.
  \label{ecparares}
\end{align}
Note that if $H_{a}=0$, the equation leads to the critical value of the scale factor, $a=a_{c}$. This is the behavior we are expecting instead of the Schwarzschild singularity.

\begin{figure}[!htb]
\centering
\includegraphics[width=0.75\columnwidth]{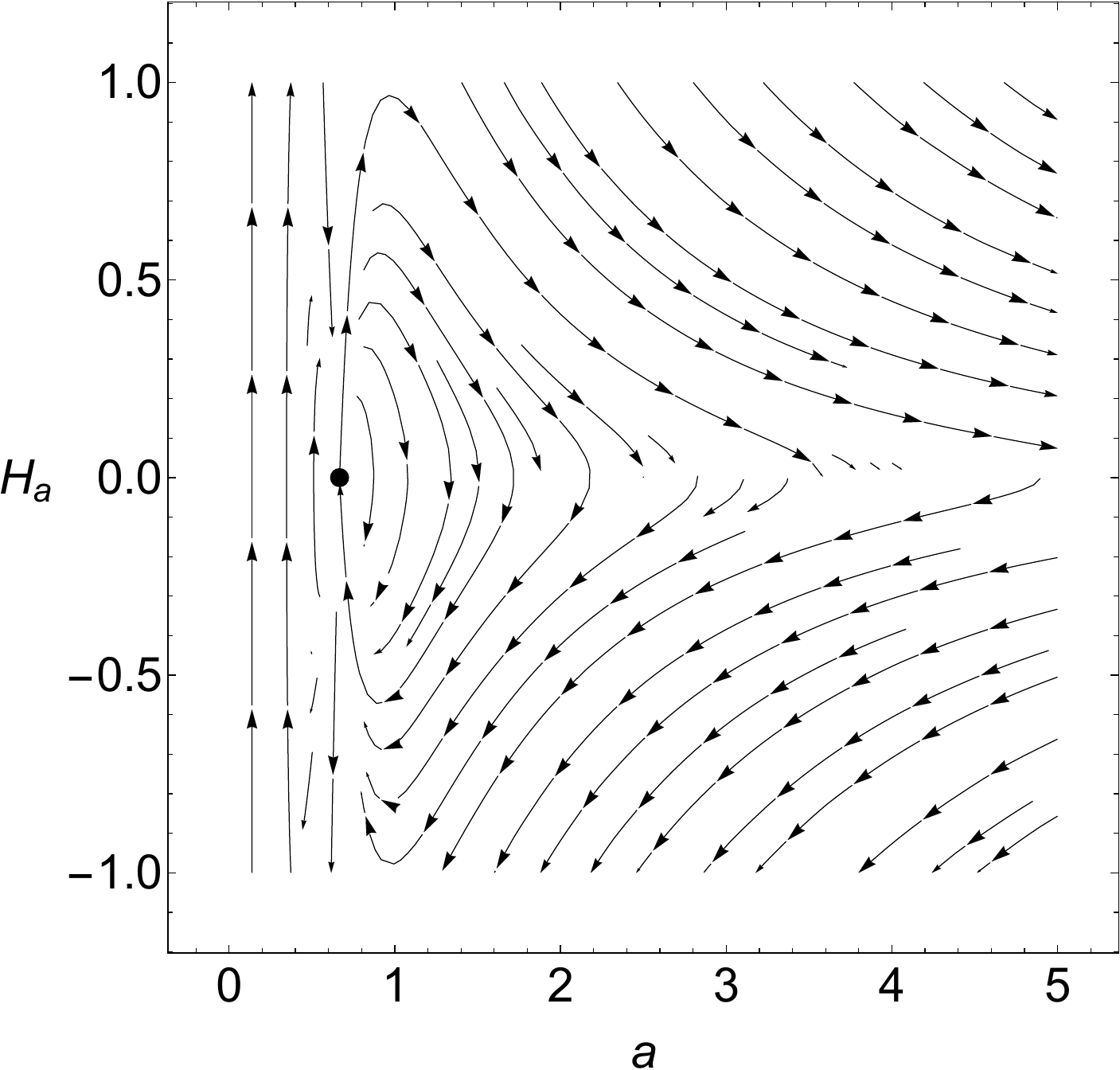} \\[2ex]
\includegraphics[width=0.75\columnwidth]{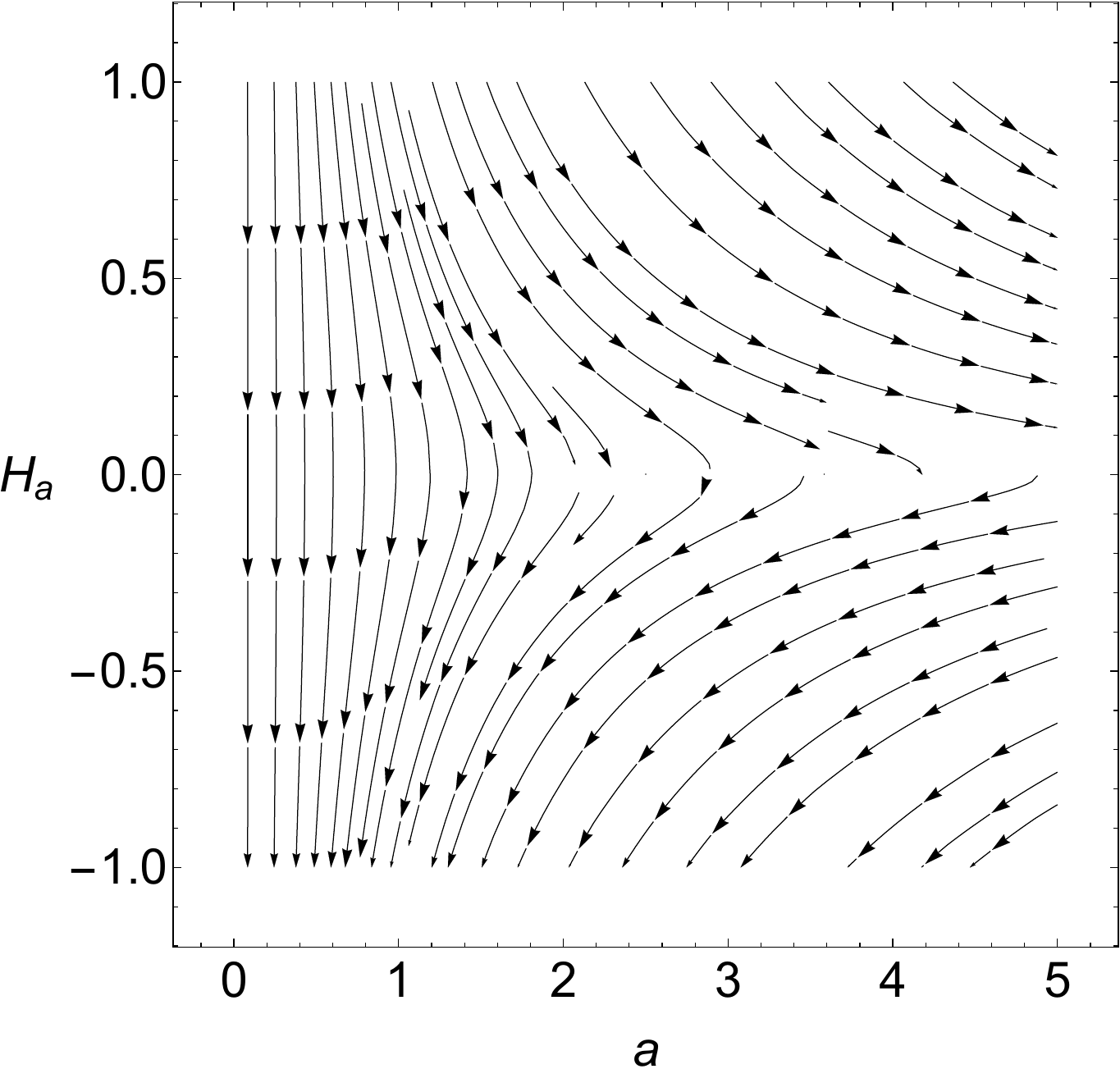}
\caption{Top panel: Phase space of Eq.~(\ref{ecparares}) when viewed as a system of ODEs in the two variables $H_a$ and $a$. The parameter $\lambda$ was set as $|\lambda|=9$. The dot represents the critical point $a_c$. Bottom panel: Phase space of Eq.~(\ref{ecparares}) in the limit $|\lambda| \rightarrow \infty$ which corresponds to standard GR. Note the inversion of the arrows near $a=0$, in contrast to the top panel.}
\label{fig:phase1}
\end{figure}

Equation~(\ref{ecparares}) can be seen as a first order ODE in the Hubble parameter or a second order ODE in the scale factor. Alternatively, one can rewrite this equation as a pair of first order ODEs in the variables $H_a$ and $a$ which then allows us to use qualitative techniques to understand the dynamics of this system. The resulting phase space diagram is shown in Fig.~\ref{fig:phase1} and demonstrates quite nicely the physical properties of this modified Schwarzschild solution.

For small $a$ one observes a divergent positive `acceleration' $\dot{H}_a$ which pushes the trajectories away from the $\dot{a}$-axis, making $a=0$ unattainable from a dynamical systems point of view, see the subsequent discussion. Moreover, the system contains a critical point with coordinates $(a,H_a)=(a_c,0)$ which replaces the classical Schwarzschild singularity. In stark contrast to the BI modified model, we note that all trajectories are attracted by the singular centre in the case of GR, i.e.~$|\lambda| \rightarrow \infty$ or $a_{c}\rightarrow 0$ (see Fig.~\ref{fig:phase1} bottom panel).

The nature of the hypersurface defined by $a=a_{c}$ can be examined by looking at Eq.~(\ref{ecparares}) in the vicinity of this critical point. In order to do this, we will put $a(t)=a_{c}(1+\varepsilon F(t))$ and keep terms of the lowest order in $\varepsilon$, while assuring a domain where $F(t)$ is bounded. We obtain
\begin{align}
  \label{ecpararesaprox}
  4\varepsilon^2(a_{c}^{-2}F^{2}-\dot{F}^{2}+F\ddot{F})+\mathcal{O}(\varepsilon^3)=0.
\end{align}
The solution of this equation is simply
\begin{align}
  \label{ecpararesaproxsol}
  F(t)=A \exp\Big(-\frac{t^{2}}{2a_{c}^{2}}+B\,t\Big),
\end{align}
where $A,B$ are integration constants. The functions $F$ and $\dot{F}$ tend to zero asymptotically in both directions of time. We can proceed now to obtain $b(t)$ with the help of~(\ref{CombiBI}), namely
\begin{align}
  \label{bpora}
  \log|b(t)| = -\frac{1}{2}\int\Big[\frac{ a^{2}-a_{c}^{2}}{ a^{4}H_{a}}+H_{a}\Big]\,dt,
\end{align}
which, in terms of $F(t)$ gives
\begin{align}
  \label{bpora2}
  \log|b(t)| = -\frac{1}{a_{c}^{2}}\int\frac{F}{\dot{F}}\,dt.
\end{align}
With the form of $F(t)$ obtained in~(\ref{ecpararesaproxsol}), the other scale factor is
\begin{align}\label{bpora3}
  b(t)=b_{0}(t-B \,a_{c}^2).
\end{align}
The constant $B$ controls the origin of time, so we can safely set $B=0$. At the end, the metric near $a_{c}$ at the lowest order is
\begin{align}
  \label{metsobresup}
  ds^{2}=dt^{2}- b_{0}^{2}\,t^2  dz^{2}- a_{c}^{2}(1+2A\,e^{-t^{2}/2a_{c}^{2}})d\Omega^{2},
\end{align}
where we have redefined $\varepsilon A \rightarrow A$. An inspection of the curvature scalars reveals

\begin{align}
  \label{escalares}
  R = -\frac{2}{a_{c}^{2}} + \xi_{1},\
  R^{(2)} = \frac{2}{a_{c}^{4}}+\xi_{2},\
  K= \frac{4}{a_{c}^{4}}+\xi_{3}
\end{align}
where $R^{(2)} = R_{\mu\nu}R^{\mu\nu}$, $K = R^{\alpha}{}_{\mu\nu\rho}R_{\alpha}{}^{\mu\nu\rho}$ and $\xi_{i}=\xi_{i}(A,t)$ $i=1,2,3$ are functions which tend to zero as $t\rightarrow\pm\infty$ for all $A$. As a matter of fact, at the critical point the metric reads, by setting $A=0$
\begin{align}
  \label{metsobreacrit}
  ds^{2}=dt^{2}-b_{0}^{2}\,t^2 dz^{2}- a_{c}^{2} d\Omega^{2},
\end{align}
and the invariants are all constant, proportional to $\lambda$ or $\lambda^{2}$ depending on the case. The two dimensional ($t,z$) part of~(\ref{metsobreacrit}) is just Minkowski space in Milne-like coordinates. For $b_{0}\neq 0$, the use of the conformal time $\tilde{t}=b_{0}^{-1}\log(t)$ and a change to Rindler coordinates ($T,X$)
\begin{align}\nonumber
  T = \exp(b_{0}\tilde{t}) \cosh(b_{0} z),\quad
  X = \exp(b_{0}\tilde{t}) \sinh(b_{0} z),
\end{align}
allows us to put the metric at $a_{c}$ in the form
\begin{align}
  \label{metsobreacritfin}
  ds^{2}=dT^{2}-dX^{2}- a_{c}^{2} d\Omega^{2}.
\end{align}
This metric has been found previously in~\cite{Gott} and corresponds to the interior of an infinitely long (in the $z$-direction) cosmic string with energy momentum tensor $\mathcal{T}_{\mu}^{\nu}=\mathrm{diag}(\rho,-p_{z},-p_{\theta},-p_{\phi})$ with $p_{\theta}=p_{\phi}=0$, $\rho=-p_{z}$, where the energy density is related to the critical radius according to
\begin{align}
  \label{densidad}
  \rho = \frac{1}{8 \pi a_{c}^{2}} = \frac{|\lambda|}{32 \pi G} \,.
\end{align}
Due to~(\ref{metsobreacritfin}) being a static metric, no vestiges remain of the cosmological character of the spacetime characterizing the region $a(t)\neq a_{c}$. On dynamical grounds, the observer approaching $a_{c}$ will experience a spacetime with bounded curvature scalars which, however, will be not attainable in finite proper time. As he/she approaches $a_{c}$, the surrounding universe turns into a static spacetime with constant positive energy density proportional to the BI constant $\lambda$ and a negative pressure along the $z$-direction. It is worth to emphasise that the energy density and pressure so obtained come from the effective (geometrical) energy-momentum tensor, and not from the introduction of any explicit matter fields through $I_{\rm matter}$.

If we view the additional terms in the field equations as matter, then we see that this matter satisfies the condition $\rho + p_z = 0$, which implies that the weak and strong energy conditions are not violated. The parameter $\lambda$ bounds the curvature scalars of the black hole interior, in such a way that the Markov's `limiting curvature hypothesis'~\cite{Markov1,Markov2,Frolov,Brand1} is dynamically and naturally implemented, without invoking junction conditions which are necessary in order to match topologically different spaces.

\section{Final comments}

Even though metric~(\ref{metsobreacritfin}) is unreachable for infalling observers, it is fair to ask what is the structure of the space beyond that asymptotic \emph{assembly space}, defined by the interior string metric. The region $a<a_{c}$ is causally disconnected from the spacetime accessible to observers in free fall. In order to study this region we assume $a$ to be small in Eq.~(\ref{ecparares}), so that it takes the simple form
\begin{align}
  \label{smalla}
  \dot{H}_a = -\frac{7}{2} H_a^2 + \frac{2}{|\lambda|a^4} \,.
\end{align}
It follows immediately that the effect of the Born-Infeld term is opposite to that of GR. This explains the inversion of the trajectories near $a=0$ in Fig.~\ref{fig:phase1}. One sees that the central Schwarzschild singularity is replaced by a singular repulsive centre which is not part of the manifold. The new term dominates for small $a$ thereby forcing the sign of $\dot{H}_a$ to be positive. As we approach the centre we are pushed away from it with larger and larger acceleration. The solution of~(\ref{smalla}) is given by $a(t)=a_0 t^{1/2}$ with $a_0^4=16/(3|\lambda|)$. Note that the units of $a_0$ are such that $a(t)$ is dimensionless, recall that $a_c$ itself has units of length. This means $a_0^2$ has units of inverse length, consistent with the result. 

One verifies that $b(t)=b_0 t^{1/2}$ is a solution of~(\ref{CombiBI}) at the same order, here $b_0$ is a constant of integration. The resulting metric near the centre is similar to the radiation dominated solution in flat FLRW cosmology and contains a strong curvature singularity at $t=0$. However, in the present context this singularity becomes repulsive and cannot be reached. Trajectories getting close to $a=0$ will repelled and will eventually approach the assembly space.

The results here obtained seem to suggest that the usual view of impending and inevitable destruction at the centre of the black hole should possibly be changed in favor of a rather more balance kind of reality.

\begin{acknowledgments}
The authors acknowledge support by The Royal Society, International Exchanges 2017, grant number \mbox{IEC\textbackslash R2\textbackslash 170013}. FF wants to thank the Department of Mathematics at UCL, where part of this investigation was undertaken. He also is indebted to Santiago Hernández for his valuable comments on dynamical systems. The authors are very grateful for valuable discussion with Cecilia Bejarano who contributed to the early stages of this project.
\end{acknowledgments}


\end{document}